# Gigantic negative magnetoresistance of nanoheterostructures described by the Fivaz model

P.V.Gorskyi

Institute of Thermoelectricity NAS and MES of Ukraine
Ukraine, Chernivtsi
gena.grim@gmail.com

*Abstract. It is shown that the negative magnetoresistance of nanoheterostructures described by the Fivaz model can become apparent not only under strong, but also under intermediate or weak degeneracy of free carrier gas in them. In so doing, in the Fivaz model it becomes apparent to a larger extent than in the case of a parabolic, though anisotropic, conduction band. The negative magnetoresistance can be both due to spin splitting and the Landau quantization proper.*

The negative longitudinal magnetoresistance of nanoheterostructures described by the Fivaz model was studied in [1]. In so doing, essential emphasis was placed on the impact of spin splitting on this effect. However, these investigations were restricted to consideration of the case of strongly open Fermi surfaces and low temperatures. In turn, in [2] it was shown that nonparabolicity as such described by the Fivaz model in case of closed Fermi surfaces results in the appearance of negative magnetoresistance area on the field dependence of electric conductivity. At the same time, the use of superlattice materials for creation of various electronic devices in a number of cases supposes the closeness of their Fermi surfaces or, moreover, the nondegeneracy of free carrier gas [3]. Therefore, of certain interest is studying of negative magnetoresistance effect under weak and intermediate degeneracy.

For this purpose, let us use general relations for the electric conductivity components obtained in [4]. In so doing, the temperature will be considered to be sufficiently high for the Shubnikov-de-Haas oscillations not to occur. Therefore, the electric conductivity of nanoheterostructure for the case when the electric and quantizing magnetic fields are perpendicular to layers is determined as:

$$\sigma_{zz} = \frac{4e^2 a^2 \rho s^2 \Delta}{3h\Xi^2 t} \left\{ \int_0^{\arccos(1-\gamma)} \varphi(x)dx + \sum_{l=1}^{\infty}(-1)^l h_l^{\sigma}\left[ \int_0^{\arccos(1-\gamma)} \varphi(x)\exp\left(l\frac{1-\cos x-\gamma}{t}\right)dx - \int_{\arccos(1-\gamma)}^{\pi} \varphi(x)\exp\left(-l\frac{1-\cos x-\gamma}{t}\right)dx \right] \right\}, \quad (1)$$

At $\gamma \leq 0$, in formula (1) only the third integral remains with integration limits from 0 to $\pi$.

In the derivation of formula (1) it was supposed that the dominant mechanism of carrier scattering is acoustic phonon deformation potential scattering, the relaxation time being inversely proportional to the product of temperature and the density of states. In formula (1), the following designations are introduced: $a$ is the distance between translation-equivalent layers of nanoheterostructure, $\rho$ is density of nanoheterostructure, $s$ is sound velocity in nanoheterostructure, $\Delta$ is halfwidth of a miniband determining carrier motion in a direction normal to layers, $\Xi$ is deformation potential constant, $\gamma = \zeta/\Delta$, $t = kT/\Delta$, $\zeta$ is chemical potential of free carrier gas. The rest of designations are generally accepted. Moreover:

$$\varphi(x) = 0.25x^{-2}\sin 2x - 0.5x^{-1}\cos 2x, \quad (2)$$

$$h_l^{\sigma} = \frac{bl/t}{\sinh(bl/t)}. \quad (3)$$

In so doing, in formula (3) $b = \mu^* B/\Delta$, $\mu^* = \mu_B m_0/m^*$, $m^*$ is electron effective mass in layer plane, $B$ is magnetic field induction, the rest of designations are generally accepted.

In the absence of nonparabolicity, formula (1) takes on the form:

$$\sigma_{zz} = \frac{8e^2 a^2 \rho s^2 \Delta}{9h\Xi^2 t}\left\{0.5\gamma[1+\mathrm{sgn}(\gamma)] + b\sum_{l=1}^{\infty}(-1)^{l-1}\frac{\exp(-lt^{-1}|\gamma|)}{\sinh(bl/t)}\right\}. \quad (4)$$

Chemical potential $\zeta$ is determined from the following equation:

$$n_0 = \frac{4m^*}{ah^2}\int_{W(x)\leq\zeta}[\zeta - W(x)]dx + \\ + \frac{eB}{\pi ah}\sum_{l=1}^{\infty}\frac{(-1)^{l-1}}{\mathrm{sh}(\mu^* Bl/kT)}\left[\int_{W(x)\leq\zeta}\exp\left(l\frac{W(x)-\zeta}{kT}\right)dx + \int_{W(x)\geq\zeta}\exp\left(l\frac{\zeta - W(x)}{kT}\right)dx\right]. \quad (5)$$

In the case of the Fivaz model, $W(x) = \Delta(1-\cos x)$ is inserted in this equation, and in the absence of nonparabolicity – $W(x) = 0.5\Delta x^2$, when $n_0$ is the bulk concentration of free carriers.

The results of calculations of the field dependences of magnetoresistance for both cases with different bulk concentrations of current carriers are given in Fig.1. In so doing, free carrier concentration is characterized by dimensionless parameter $\gamma_0 = \zeta_0/\Delta$, where $\gamma_0$ is chemical potential of free carrier gas with the absolute zero temperature.

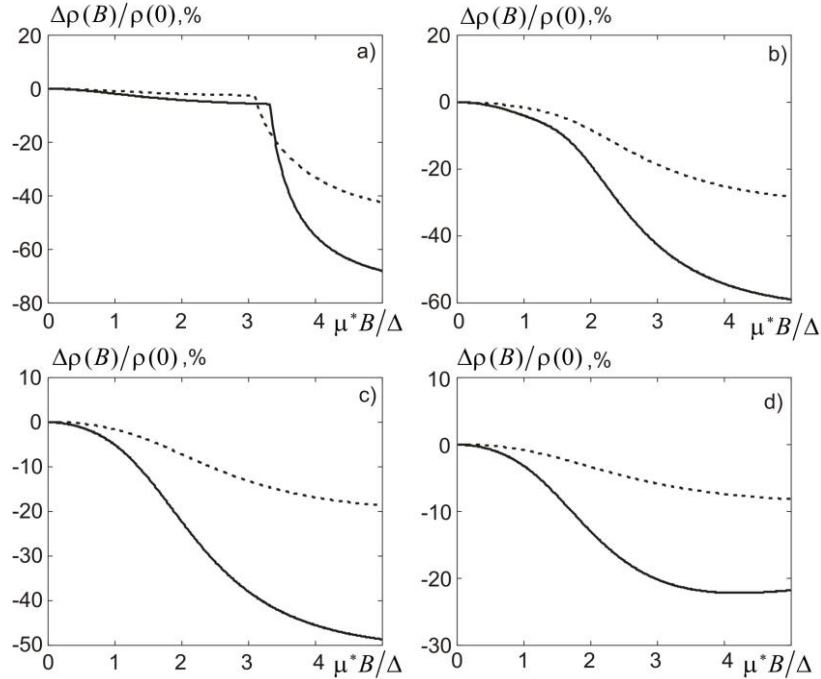

*Fig.1. Field dependences of magnetoresistance at $kT/\Delta = 1$ и: a) $\gamma_0 = 0.5$; b) $\gamma_0 = 1$; c) $\gamma_0 = 1.5$; d) $\gamma_0 = 2$. Continuous curves correspond to the Fivaz model, dashed curves – to absence of nonparabolicity*

From the figure it is seen that longitudinal magnetoresistance of nanoheterostructures is mostly negative, and in the case of the Fivaz model it is much more pronounced that in the absence of nonparabolicity. The very presence of negative magnetoresistance can be explained as follows. Eq. (5) implies that in the absence of oscillations the chemical potential of carrier gas is increased. And as long as magnetic field growth to a certain extent is equivalent to crystal cooling, the first members of Eqs. (5) and (4) that are increasing functions of magnetic field are predominant. Therefore, it is clear that until these terms are essential, the magnetoresistance will be mostly negative. However, the sum of the second and third terms, generally speaking, is a nonmonotonic function of a magnetic field. Such situation occurs because, on the one hand, the sign of explicitly magnetic field-dependent addition to electric conductivity is determined by two competent processes of carrier thermal umklapp: from the lower Landau subbands to chemical potential level and from chemical potential level to the upper Landau subbands. If the former process dominates, the contribution of this addition to magnetoresistance is positive, and if the latter process dominates, it is negative. However, the value of this contribution almost exponentially decays with increasing ratio of the distance between the Landau subbands to thermal motion energy. Besides, in case of nonparabolicity described by the Fivaz model the negative magnetoresistance is more pronounced due to restriction of carrier scattering according to energies and a nonmonotonic change in squared velocity of carriers as the function of energy. However, at $\gamma_0 = 2$, in the case of the Fivaz model, after intersection of miniband ceiling by chemical potential level there is inversion of magnetoresistance, and it becomes weakly positive, while in the absence of nonparabolicity it remains negative. On the whole, in the considered range of magnetic fields and carrier concentrations, in the case of the Fivaz model, the negative magnetoresistance reaches 24-67%, whereas in the absence of nonparabolicity it does not exceed 9-42%. Therefore, gigantic negative longitudinal magnetoresistance can be considered as a diagnostic sign of the Fivaz model applicability to nanoheterostructures.